\documentclass[aps,pra,reprint,onecolumn,amsmath,amssymb,amsfonts,preprintnumbers,showkeys,showpacs]{revtex4-1}
\usepackage{bm,graphicx,xcolor,microtype}
\newcommand{\md}{\mathcal{D}}
\newcommand{\mR}{\mathcal{R}}

\newcommand{\alert}[1]{\textcolor{black}{#1}}

\begin{document}

\title{Exact wave functions for concentric two-electron systems}

\author{Pierre-Fran\c{c}ois Loos}
\email{loos@rsc.anu.edu.au}
\author{Peter M. W. Gill}
\email{peter.gill@anu.edu.au}
\affiliation{Research School of Chemistry, 
Australian National University, Canberra, ACT 0200, Australia}

\begin{abstract}
We show that the exact solution of the Schr\"odinger equation for two electrons confined to two distinct concentric rings or spheres can be found in closed form for particular values of the ring or sphere radii. \alert{In the case of rings, we report exact polynomial and irrational solutions. In the case of spheres, we report exact polynomial solutions for the ground and excited states of $S$ symmetry.}
\end{abstract}

\maketitle

\section{Introduction}

\alert{Exactly or quasi-exactly \footnote{\alert{Quasi-exact solvability means that part, but not all, of the energy spectrum can be solved explicitly.}} solvable models \cite{Ushveridze} can be seen as ``theoretical laboratory'' for physicists and chemists \cite{EbrahimiFard12}, due to their usefulness for both testing and developing approximate methods, such as density-functional \cite{ParrBook, Filippi94, UEGs12}, GW \cite{Schindlmayr13}, and density-matrix functional \cite{BenavidesRiveros12} theories. Indeed, understanding correlation effects remains one of the central problem in theoretical quantum chemistry and physics and is the main goal of most of the new theories and models in this research area \cite{Helgaker, ParrBook}.}


In a recent series of papers \cite{QuasiExact09, LoosExcitSph, QR12}, we have found exact solutions for several new quasi-exactly solvable models. In particular, we have shown that one can solve the Schr\"odinger equation for two electrons confined to a ring \cite{QR12} or to the surface of a sphere \cite{QuasiExact09, LoosExcitSph} for particular values of the ring or sphere radius. \alert{In particular, we showed that some of the solutions exhibit the Berry phase phenomenon, i.e. if one of the electrons moves once around the ring and returns to its starting point, the wave function of the system changes sign \cite{QR12}.} Subsequently, Guo \textit{et al.}~reported some exact solutions for two vertically coupled quantum rings using the same technique \cite{Guo12} but their analysis is incomplete \footnote{Although the present study focuses on different systems, our analysis can be applied to find additional exact solutions for their system.}.

\alert{From the experimental point of view, much effort has been devoted in the past decade} to the fabrication of GaAs/AlGaAs  concentric double quantum rings using droplet-epitaxial techniques \cite{Kuroda05, Mano05, Muhle07} in order to study the Aharonov-Bohm effect \cite{Aharonov59} and the influence of the Coulomb interaction on the magnetic properties \cite{Muhle07}. More recently, accurate theoretical calculations, using local spin density functional theory \cite{Malet06}, exact diagonalization \cite{Szafran05} and quantum Monte Carlo \cite{Colletti09} techniques, have shown that Wigner molecules \cite{Wigner34} are formed at low density.

Multiple-shell fullerenes or ``buckyonions'' \cite{Ugarte92, Ugarte93} are another kind of concentric material which have attracted research attention for many years. Since the 1985 discovery of fullerene \cite{Kroto85}, experimental search for such carbon-based materials have been very fruitful, leading to the discovery of carbon nanotubes, boron buckyballs, etc. The morphology of buckyonions and the interaction energy between shells have been investigated using first principles calculations \cite{Lu94, Bates98} and continuum models \cite{Guerin97, Baowana07}.

\alert{A minimalist model for these systems consists of electrons confined to concentric rings or spheres in three-dimensional space}. In Ref. \cite{LoosConcentric}, we reported a comprehensive numerical study of the singlet ground state of two electrons on concentric spheres with different radii. We analyzed the strengths and weaknesses of several electronic structure models, ranging from the mean-field approximation (restricted and unrestricted Hartree-Fock solutions) to configuration interaction expansion, leading to near-exact wave functions and energies. Berry and collaborators also considered this model to simulate doubly-excited states of helium  \cite{Ezra83} and the rovibrational spectra of the water molecule in both the ground \cite{Natanson84} and excited states \cite{Natanson86}. More recently, the model has been applied to quantum-mechanical calculations of large-amplitude light atom dynamics in polyatomic hydrides \cite{Deskevich05, Deskevich08}, confirming again the broad applicability of theoretical models of concentric electrons. 

In this Article, we extend our analysis to the case where two electrons are located on concentric rings or spheres with different radii, focusing our attention on the rotationally-invariant (i.e. $S$) states. \alert{Although our principal concerns are concentric rings (i.e. $\md=1$) and concentric spheres (i.e. $\md=2$), we consider the general case in which the electrons are confined to concentric $\md$-spheres of radius $R_1$ and $R_2$ \footnote{A $\md$-sphere is the surface of a ($\md+1$)-dimensional ball}}. For $R_1 \neq R_2$, the singlet and triplet states are degenerate and one can easily switch from singlets to triplets (or vice versa) by antisymmetrizing (or symmetrizing) the spatial wave function \cite{LoosConcentric}. 

\alert{In Sec.~\ref{sec:hamiltonian}, we define the Hamiltonian of the system and various other important quantities. In Sec.~\ref{sec:HF}, we give the explicit expression of the Hartree-Fock (HF) wave function and energy for two electrons on concentric rings or spheres. The general method to obtain exact solutions of the Sch\"odinger equation is given in Sec.~\ref{sec:exact}. Finally, we report the explicit expression of the exact wave functions in Sec.~\ref{sec:results}.}

\section{\label{sec:hamiltonian}Hamitonian}

\begin{figure}
\includegraphics[width=0.4\textwidth]{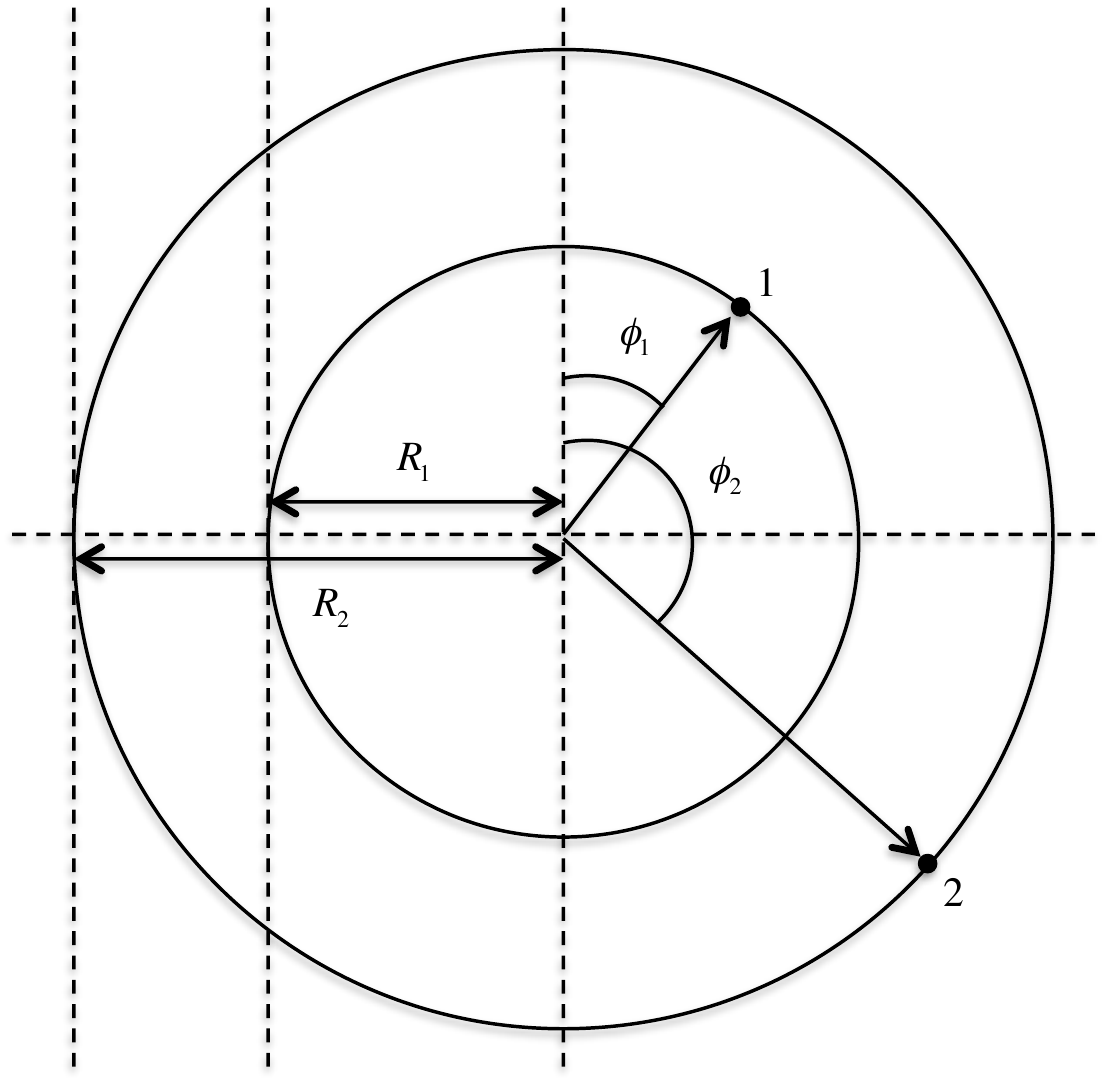}
\caption{
\label{fig:fig1}
\alert{Two electrons on concentric rings: electron 1 is located on the inner ring of radius $R_1$ and electron 2 is located on the outer ring of radius $R_2$. $\phi_1$ and $\phi_2$ are the azimuthal angles of electrons 1 and 2, respectively.}}
\end{figure}

\alert{The model consists of two concentric rings or spheres of radii $R_1$ and $R_2$, each bearing one electron, and, in the remainder of this study, we assume $R_1 < R_2$ (see Fig.~\ref{fig:fig1}). The electronic Hamiltonian of the system is
\begin{equation}
\label{H}
	\Hat{H} = \Hat{T} + u^{-1},
\end{equation}
where 
\begin{equation}
\label{T}
	\Hat{T} = \Hat{T}_1 + \Hat{T}_2 =  -\frac{\nabla_1^2}{2R_1^2} - \frac{\nabla_2^2}{2R_2^2} 
\end{equation}
is the kinetic energy operator and $u^{-1}$ is the Coulomb operator. In \eqref{T}, $\nabla_i^2$ is the angular part of the Laplacian for the $i$th electron, and reads
\begin{equation}
	\nabla_i^2 = 
	\begin{cases}
		\frac{\partial^2}{\partial\phi_i^2},	&	\text{rings ($\md=1$)},
		\\
		\frac{\partial^2}{\partial\theta_i^2} + \cot \theta_i \frac{\partial}{\partial\theta_i} + \frac{1}{\sin^2 \theta_i} \frac{\partial^2}{\partial\phi_i^2},	&	\text{spheres ($\md=2$)},
	\end{cases}
\end{equation}
where $\phi_i \in [0,2\pi)$ and $\theta_i \in [0,\pi]$ are the usual azimuthal and polar angles of the $i$th electron. The interelectronic distance
\begin{equation}
\label{u}
	u=\sqrt{R_1^2+R_2^2-2R_1R_2\cos \omega}
\end{equation}
ranges from $R_2-R_1$ to $R_1+R_2$, and $\omega \in [0,\pi]$ is the interelectronic angle and satisfies
\begin{equation}
	\cos \omega = 
	\begin{cases}
		\cos(\phi_1-\phi_2),	&	\text{rings ($\md=1$)},
		\\
		\cos \theta_1\cos \theta_2 + \sin \theta_1 \sin \theta_2 \cos(\phi_1 - \phi_2),	&	\text{spheres ($\md=2$)}.
	\end{cases}
\end{equation}}
For notational simplicity, we introduce
\begin{align}
	\frac{1}{\sigma} = \frac{1}{\mR^2} & = \frac{1}{2R_1^2} + \frac{1}{2R_2^2},
	&
	\sigma_{\pm} & = \mR_{\pm}^2 = (R_1 \pm R_2)^2,
\end{align}
where $\mR$ is an ``effective'' radius, and $\mR_{-}$ and $\mR_{+}$ are respectively the lowest and highest possible values of $u$.

\section{\label{sec:HF}Hartree-Fock approximation}

\alert{Within the HF approximation \cite{SzaboBook}, the high symmetry of the system implies that there is a symmetric solution (called SHF) in which each orbital is constant over its ring/sphere. For two electrons on concentric rings ($\md=1$), the resulting wave function and energy are
\begin{align}
	\Psi_\text{SHF} & = \frac{1}{2\pi \sqrt{R_1 R_2}},
	&
	E_\text{SHF} & = \frac{2}{\pi(R_2-R_1)} K\left( -\frac{4R_1R_2}{\sigma_{-}} \right),
\end{align}
where $K(x)$ is the complete elliptic integral of the first kind \cite{NISTbook}. We note that, for $R_1=R_2$, $E_\text{SHF}$ diverges due to the singularity of the Coulomb operator.} 

\alert{For two electrons on concentric spheres ($\md=2$), the SHF wave function and energy are \cite{LoosConcentric}
\begin{align}
	\Psi_\text{SHF} & = \frac{1}{4\pi R_1 R_2},
	&
	E_\text{SHF} & = \frac{1}{R_2},
\end{align}
and we note that $E_\text{HF}$ only depends on the radius of the second electron.}

\alert{For certain values of the radii, a second, lower-energy HF solution exists, in which the two electrons tend to localize on opposite sides of the rings/spheres \cite{LoosConcentric, QR12}.}

\section{\label{sec:exact}Exact solutions}
For states of zero angular momentum ($S$ states), the wave function depends only on $u$ and, \alert{using \eqref{u}, the Hamiltonian \eqref{H} can be recast}
\begin{equation}
	\Hat{H} 
	= \frac{1}{4\sigma} \Bigg\{ \left[ u^2 + \frac{\sigma_{-} \sigma_{+}}{u^2} - (\sigma_{-}+\sigma_{+})\right]  \frac{d^2}{du^2} 
	+ \left[(2\md-1)u - \frac{(\md-1)(\sigma_{-}+\sigma_{+})}{u} - \frac{\sigma_{-}\sigma_{+}}{u^3} \right] \frac{d}{du} \Bigg\}
	+ \frac{1}{u}.
\end{equation}
The resulting Schr\"odinger equation is Fuchsian \cite{NISTbook} with singularities at $-\mR_{+}$, $-\mR_{-}$, $0$, $\mR_{-}$ and $\mR_{+}$, \alert{which means that it can be written in the form \footnote{\alert{See Ref.~\cite{NISTbook} p. 718 for more details about the Fuchian equation.}}}
\begin{equation}
	\label{Fuchsian}
	\Psi^{\prime\prime}(u) + \left[ \frac{\md/2}{u+\mR_{+}} + \frac{\md/2}{u+\mR_{-}} - \frac{1}{u} + \frac{\md/2}{u-\mR_{-}} + \frac{\md/2}{u-\mR_{+}} \right] \Psi^{\prime}(u)
	+ \frac{V(u)}{(u-\mR_{-}^2)(u-\mR_{+}^2)} \Psi(u) = 0,
\end{equation}
where $V(u)=4\mR^2 u(1-E u)$ is a Van Vleck polynomial \alert{\footnote{\alert{See Ref.~\cite{NISTbook} p. 718 for more details about Van Vleck polynomials.}}.} The Fuchsian equation is a generalization of the Heun equation \cite{Ronveaux} (which appears when $R_1=R_2$ \cite{QuasiExact09, LoosExcitSph, QR12}) for higher number of singularities \alert{\footnote{\alert{See Ref.~\cite{NISTbook} p. 710 for more details about the Heun equation.}}}. 

Knowing that Eq.~\eqref{Fuchsian} is Fuchsian, we seek solutions of the form 
\begin{equation}
\label{sol}
	\Psi_{n,m}^{(a,b)}(u) = \left(1-\frac{u^2}{\mR_{+}^2}\right)^{a/2} \left(1-\frac{u^2}{\mR_{-}^2}\right)^{b/2} S_{n,m}^{(a,b)}(u)
\end{equation}
where $(a,b)$ can take the values $(0,0)$, $(2-\md,0)$, $(0,2-\md)$ and $(2-\md,2-\md)$. The functions 
\begin{equation}
\label{Sabu}
	S_{n,m}^{(a,b)}(u)= \sum_{k=0}^{n} s_k u^k
\end{equation}
are called $n$th-degree Stieltjes polynomials \alert{\footnote{\alert{See Ref.~\cite{NISTbook} p. 719 for more details about Stieltjes polynomials.}}}.  The index $m$ is the number of nodes between $\mR_{-}$ and $\mR_{+}$.  The different possible values for the set $(a,b)$ are determined by ensuring that the substitution of \eqref{sol} into \eqref{Fuchsian} conserves the Fuchsian nature of the differential equation and the Van Vleck nature of the polynomial $V(u)$.  After making this substitution, we find that $S_{n,m}^{(a,b)}$ satisfies
\begin{equation}
\label{H-S}
	\frac{1}{4\sigma} \left[ P(u) \frac{d^2S_{n,m}^{(a,b)}(u)}{du^2} + Q^{(a,b)}(u) \frac{dS_{n,m}^{(a,b)}(u)}{du} \right] + \frac{S_{n,m}^{(a,b)}(u)}{u} = E^{(a,b)} S_{n,m}^{(a,b)}(u),
\end{equation}
with
\begin{align}
	P(u) & = u^2 + \frac{\sigma_{-} \sigma_{+}}{u^2} - (\sigma_{-}+\sigma_{+}),
	\\
	Q^{(a,b)}(u) & =  \left[(2\md-1)+2(a+b)\right]u - \frac{(\md-1)(\sigma_{-}+\sigma_{+})+2(a\,\sigma_{-}+b\,\sigma_{+})}{u} - \frac{\sigma_{-}\sigma_{+}}{u^3},
	\\
	E^{(a,b)} & = E +
	\begin{cases}
		0,					&	(a,b)=(0,0),
		\\
		\md(\md-2)/(4\sigma),	&	(a,b)=(2-\md,0) \text{ or } (a,b)=(0,2-\md),
		\\
		(\md-2)/\sigma,			&	(a,b)=(2-\md,2-\md).
	\end{cases}
\end{align}

\alert{Substituting \eqref{Sabu} into \eqref{H-S},} it can be shown that the coefficients $s_k$ satisfy the five-term recurrence relation
\begin{multline}
\label{recurrence}
	s_{k+4} = \frac{1}{\sigma_{-} \sigma_{+}} \frac{1}{k+4} 
	\Bigg\{ \frac{(k+\md)(\sigma_{-}+\sigma_{+})+2(a\,\sigma_{-}+b\,\sigma_{+})}{4\sigma} s_{k+2} 
	\\
	- \frac{4\sigma}{k+2} \left[ s_{k+1} + \left( \frac{k[k+2\md+2(a+b-1)]}{4\sigma} - E^{(a,b)}\right)s_{k} \right] \Bigg\},
\end{multline}
with the starting values
\begin{align}
	s_0 & =1,	&	s_1 & = 0,	&	s_3 & = - \frac{4\sigma}{3\sigma_{-}\sigma_{+}}.
\end{align}
One can note the similarity of the three-term recurrence relation obtained for $R_1=R_2$ \cite{QuasiExact09, LoosExcitSph, QR12} and the term in square brackets in \eqref{recurrence}. The condition $s_1=0$ shows that the exact wave functions do not contain any term proportional to the interelectronic distance $u$. Indeed, due to the spatial separation of the electrons, there is no need to satisfy the electron-electron Kato cusp condition \cite{Kato57}.

To define the $n$th-degree polynomial $S_{n,m}^{(a,b)}$ completely, we need to find the values of the inner and outer radii $R_1$ and $R_2$, the exact energy $E$, and the coefficient $s_2$. These are obtained by solving $s_{n+1} = s_{n+2} = s_{n+3} = s_{n+4}=0$, which are functions of these four unknown quantities. For $s_{n+1} = s_{n+2} = s_{n+3}=0$, the condition 
\begin{equation}
	E^{(a,b)} = \frac{n[n+2\md+2(a+b-1)]}{4\sigma}
\end{equation}
ensures that $s_{n+4} = 0$ and allows us to determine the exact energy knowing the degree of the polynomials and the effective radius.

\begin{table}
\caption{
\label{tab:rings}
\alert{Exact wave functions for ground and excited states of two electrons on concentric rings}.
}
\begin{ruledtabular}
\begin{tabular}{cccccc}
Number ($m$)		&	Degree ($n$)	&	Family		&	Radius			&	Radius				&	Energy	\\	
of nodes			&	of polynomial	&	$(a,b)$		&	$R_1$			&	$R_2$				&	$E$	\\	
\hline
0	&	4	&	$(0,0)$		&	6.56723			&	15.0297				&	0.0552268	\\	
0	&	5	&	$(0,0)$		&	9.69674			&	14.7302				&	0.0476376	\\	
0	&	6	&	$(0,0)$		&	14.5140			&	18.9048				&	0.0339531	\\	
0	&	7	&	$(0,0)$		&	20.4922			&	24.6415				&	0.0246731	\\	
0	&	8	&	$(0,0)$		&	27.5378			&	31.5687				&	0.0185769	\\	
1	&	5	&	$(1,0)$		&	17.2072			&	47.7424				&	0.0171724	\\	
1	&	6	&	$(1,0)$		&	20.3910			&	37.9387				&	0.0189862	\\	
1	&	7	&	$(1,0)$		&	26.3702			&	40.6719				&	0.0163405	\\
1	&	8	&	$(1,0)$		&	33.8677			&	46.8166				&	0.0134467	\\
1	&	4	&	$(0,1)$		&	8.10699			&	19.293				&	0.0559434	\\	
1	&	5	&	$(0,1)$		&	10.9336			&	16.1063				&	0.0549899	\\	
1	&	6	&	$(0,1)$		&	16.0422			&	20.4823 				&	0.0384000	\\	
1	&	7	&	$(0,1)$		&	22.3238			&	26.4996 				&	0.0274453	\\	
1	&	8	&	$(0,1)$		&	29.6705			&	33.7181 				&	0.0204070	\\	
2	&	7	&	$(0,0)$		&	12.1683			&	17.5157				&	0.0613308	\\			
2	&	8	&	$(0,0)$		&	17.5501			&	22.0442				&	0.0424363	\\	
2	&	5	&	$(1,1)$		&	19.9728			&	57.0944				&	0.0172333	\\	
2	&	6	&	$(1,1)$		&	22.2550			&	40.4332				&	0.0210458	\\	
2	&	7	&	$(1,1)$		&	28.4414			&	42.9590				&	0.0180032	\\	
2	&	7	&	$(1,1)$		&	36.2331			&	49.2980				&	0.0146648	\\			
3	&	7	&	$(1,0)$		&	24.9101			&	81.0762				&	0.0141096	\\			
3	&	8	&	$(1,0)$		&	24.1637			&	43.1663				&	0.0227745	\\
4	&	8	&	$(1,1)$		&	26.1359			&	46.2032				&	0.0241549	\\	
\end{tabular}
\end{ruledtabular}
\end{table}

\section{\label{sec:results}Results and discussion
}
\subsection{Concentric rings}
\alert{For two electrons on rings of radius $R_1$ and $R_2$ (i.e. $\md=1$), we have four different families of solutions corresponding to the values $(a,b)=(0,0)$, $(1,0)$, $(0,1)$ and $(1,1)$.} They correspond to the wave functions
\begin{align}
	\Psi_{n,m}^{(0,0)}(u) & = S_{n,m}^{(0,0)}(u),	&	\Psi_{n,m}^{(1,0)}(u) & = \cos \frac{\omega}{2} S_{n,m}^{(1,0)}(u),
	\\
	\Psi_{n,m}^{(0,1)}(u) & = \sin \frac{\omega}{2} S_{n,m}^{(0,1)}(u),	&	\Psi_{n,m}^{(1,1)}(u) & = \sin \omega \,S_{n,m}^{(1,1)}(u).
\end{align}
 As shown in Table \ref{tab:rings}, the $(0,0)$ family produces ground states and excited states wave functions, while the other families produce only excited states with at least one node ($m \ge 1$) for $(1,0)$ and $(0,1)$ and two nodes ($m \ge 2$) for $(1,1)$.

The first nodeless solution is
\begin{equation}
	S_{4,0}^{(0,0)}(u)=1 - \frac{49}{1521}  u^2 - \frac{343}{118638} u^3 - \frac{16807}{240597864}  u^4,
\end{equation}
for
\begin{align}
	R_1 & = \frac{13}{7} \sqrt{3 (13 - \sqrt{78})},
	&
	R_2 & = \frac{13}{7} \sqrt{3 (13 + \sqrt{78})},
\end{align}
and the energy of $E=28/507=0.0552268$.

The first solution which has a single node at $\omega=0$ is produced by the family $(0,1)$ for $n=4$ and reads
\begin{equation}
	S_{4,1}^{(0,1)}(u)=1 - \frac{1250}{58653} u^2 - \frac{15625}{9853704} u^3 - \frac{390625}{12231731232} u^4,
\end{equation}
for
\begin{align}
	R_1 & = \frac{7}{25} \sqrt{399 (7 - 2\sqrt{6})},
	&
	R_2 & = \frac{7}{25} \sqrt{399 (7 + 2\sqrt{6})},
\end{align}
and the energy of $E=625/11172=0.0559434$.

\begin{table}
\caption{
\label{tab:spheres}
\alert{Exact wave functions  for ground and excited states of two electrons on concentric spheres}.
}
\begin{ruledtabular}
\begin{tabular}{cccccc}
Number ($m$)		&	Degree ($n$)	&	Family		&	Radius			&	Radius				&	Energy	\\	
of nodes			&	of polynomial	&	$(a,b)$		&	$R_1$			&	$R_2$				&	$E$	\\	
\hline
0	&	5	&	$(0,0)$		&	13.4767			&	27.3109				&	0.0299540	\\	
0	&	6	&	$(0,0)$		&	18.0008			&	27.7797				&	0.0262917	\\	
0	&	7	&	$(0,0)$		&	24.2569			&	32.8421				&	0.0206850	\\			
0	&	8	&	$(0,0)$		&	31.7451			&	39.7970				&	0.0162370	\\	
1	&	7	&	$(0,0)$		&	17.0759			&	34.7323				&	0.0335356	\\	
1	&	8	&	$(0,0)$		&	21.3402			&	31.6212				&	0.0319596	\\	
\end{tabular}
\end{ruledtabular}
\end{table}

\subsection{Concentric spheres}
In the case of two electrons on concentric spheres (i.e. $\md=2$), there is only one family of solutions and it is associated with $(a,b) = (0,0)$. The first few exact solutions are reported in Table \ref{tab:spheres}.
The first solution is obtained for $n=5$ and is a ground state wave function 
\begin{equation} \label{eq:S5000}
	S_{5,0}^{(0,0)}(u) = 1 - 0.0161923\,u^2 - 0.0012233\,u^3 - 0.000033429\,u^4 - 3.14705\times10^{-7}\,u^5,
\end{equation}
associated with the energy $E = 0.029954$ and the radii $R_1=13.4767$ and $R_2=27.3109$.

Wave function terms proportional to $u$ are difficult to model using configuration interaction (CI) methods, and it is common to use explicitly correlated methods \cite{Kutzelnigg85} to improve the convergence of such calculations \cite{Helgaker}.  This was well illustrated in our $R_1 = R_2$ calculations \cite{TEOAS09}, where we observed slow convergence of the CI expansion with respect to basis set size.  In contrast, the absence of a linear term in the wave function \eqref{eq:S5000} explains the rapid convergence of our $R_1 \neq R_2$ calculations \cite{LoosConcentric}.

\section{Conclusion}
\alert{In this Article, we have shown that the Schr\"odinger equation for two electrons confined to concentric rings and spheres can be solved exactly for particular sets of the radii. In the case of concentric rings ($\md=1$), we have found four families of exact wave functions. In this case of two concentric spheres ($\md=2$), we have found only one family.}

We have not reported any exact solutions for $\md \ge 3$ in the present manuscript, but these are not difficult to derive. However, we note that only the $(a,b) = (0,0)$ family is physically significant because, for $(a,b) \neq (0,0)$, the exact solutions of \eqref{H-S} are divergent at $u=0$ or $u=\mR_{+}$.

A natural extension of the system studied in this Article would be to consider two rings of different radii vertically separated by a distance $d$. We believe this system is also quasi-exactly solvable, but the derivation would be more complicated because of the larger number of singularities in the Fuchsian equation. For this reason, we chose not to consider this system in the present study.

\begin{acknowledgements}
PMWG thanks the Australian Research Council (Grants DP0984806, DP1094170, and DP120104740) for funding. PFL thanks the Australian Research Council for a Discovery Early Career Researcher Award (Grant DE130101441).
\end{acknowledgements}

\end{document}